\magnification=1200
\baselineskip=18truept


\def\draftversion{N}

\if \draftversion Y


\fi

\def\plbfirst{1}
\def\npblong{2}
\def\exactlymassless{3}
\def\trunca{4}
\def\prl{5}
\def\scri{6}
\def\multiplemass{7}
\def\matrixbook{8}
\def\lanczos{9}
\def\rebbi{10}
\def\higham{11}
\def\borici{12}
\def\twod{13}
\def\twchiu{14}

\line{\hfill RU--98--49}
\vskip 2cm
\centerline {\bf Minimizing storage in implementations of 
the overlap lattice-Dirac operator.}
\vskip 1cm
\centerline{Herbert Neuberger}
\vskip .25cm
\centerline{\tt neuberg@physics.rutgers.edu}
\vskip 1.5cm
\centerline{\it Department of Physics and Astronomy}
\centerline{\it Rutgers University}
\centerline{\it Piscataway, NJ 08855--0849}
\vskip 2cm
\centerline{\bf Abstract}
\vskip .5cm
The overlap lattice-Dirac operator contains the sign function
$\epsilon (H)$. Recent practical implementations replace
$\epsilon (H)$ by a ratio of polynomials, $H P_n (H^2 )/Q_n (H^2)$,
and require storage of $2n+2$ large vectors. 
Here I show that one can use only 4 large vectors
at the cost of executing the core conjugate algorithm twice. 
The slow-down might be less than by a factor of 2, depending
on the architecture of the computer one uses. 

\vfill\eject

The overlap [\plbfirst, \npblong] has produced a new lattice
Dirac operator [\exactlymassless, \trunca]. Using this operator
one can define QCD on the lattice with explicit 
exact global chiral symmetries. Full, exact chiral symmetry
on the lattice without fine tuning could not be achieved
with previous lattice Dirac operators.
The new operator is given by:

$$
D={{1+\gamma_5 \epsilon (H)}\over 2},\eqno{(1)}$$
where $H=\gamma_5 D^\prime$. There is some freedom
in choosing $D^\prime$. The simplest choice is 
$D^\prime = D_W$. $D_W$, the Wilson-Dirac operator,
in $d\ge 3$-dimensions, is taken with hopping parameter $\kappa$ 
close to the middle of the allowed 
range $({1\over{2d}},{1\over{2d-4}})$. 
Choosing $D^\prime =D_W$
maximizes the sparseness of $H$ and thus minimizes 
the cost of evaluating $Hb$ for generic vectors $b$. 

Recent implementations of the overlap lattice Dirac operator
[\prl, \scri] 
employ a sum of terms of the form ${1\over {H^2 +\sigma^s}}$
with $s=1,..,n$ and $0<\sigma^1<\sigma^2...<\sigma^n$:
$$
\epsilon(H)=\lim_{n\to\infty}{H\over n} \sum_{s=1}^n
w^s {1\over{H^2 +\sigma^s}}.\eqno{(2)}$$ 
Using the method of multiple shifts [\multiplemass], the
number of $H^2 p$ operations required
to evaluate the action of the right hand side of equation
(2) on vectors $p$ for arbitrary finite $n$ becomes independent of $n$. 
Actually, the number of operations is not much more than
required to compute ${1\over{H^2 +\sigma^1}} b$ by
the simple conjugate gradient (CG) 
algorithm [\matrixbook]. This saving comes at some 
expense: one needs to store $2n+2$ vectors.

The main objective of this paper is to suggest a way to avoid 
storing $2n+2$ vectors. For large lattices this might
be necessary. The saving in
storage comes at the price of executing 
two passes over the basic CG algorithm 
instead of a single pass. Thus,
at the cost of doubling the amount of $H^2 p$ operations, 
storage also becomes almost $n$ 
independent, and does not exceed
substantially the 
requirements of computing ${1\over{H^2 +\sigma^1}} b$ by CG.
The basic trick is similar to the way large storage demands
are avoided if one uses the Lanczos method [\lanczos] to
compute not only eigenvalues but also a few eigenvectors.

Below, I shall first review the single-pass method following 
the notations and methods of Jegerlehner [\multiplemass]. 
Next, I construct the two-pass
method. I end discussing possible
advantages of using the two-pass implementation. 

For large enough $n$, typically, $\sigma^1$ is so small
relative to the lowest eigenvalue of $H^2$ that nothing is lost
by driving the algorithm with the inversion of $H^2$ itself,
rather than $H^2+\sigma^1$. In any case, the algorithm is easily
altered to make $H^2 +\sigma^1$ the driver. Below, I use just
$H^2$. I assume that a vector $b$ is given and my objective is to compute
$\epsilon(H)b$. 

\vskip .5cm
{\bf Pseudocode for the single-pass version.}
\vskip .3cm
\noindent Input vector: $b$.

\noindent Vector variables: $w$, $r$, $p$, $x^s$, $p^s$. 

\noindent Scalar variables: $\alpha$, $\beta$, $\zeta^s$, $\rho$,
${\rm tolerance}$, $b_{\rm norm}$.

\vskip .2cm
\noindent{\it Initialize}: $r_0=b$, $p_0=b$, $\rho_0 =(b,b)$, $b_{\rm norm}=
\sqrt{\rho_0}$, $\alpha_0 =0$, $\beta_{-1} =1$,  
for $s=1,2,..,n$ $\{ x_0^s=0,~ p_0^s =b,
~\zeta_{-1}^s=1,~\zeta_0^s =1 \}$.
\vskip .2cm
\noindent{\it Iterate}: for $i=0,1,2,3,...$:
\vskip .1cm
\noindent $\{$
$$\eqalign{
w_i &= H^2 p_i \cr
\beta_i &=-\rho_i / (p_i, w_i )\cr}$$

{\bf \romannumeral1.} for $s=1,2,...n$:

$\{$
$$
\zeta_{i+1}^s={{\zeta_i^s \zeta_{i-1}^s \beta_{i-1}}\over
{\beta_i\alpha_i (\zeta_{i-1}^s - \zeta_i^s ) + \zeta_{i-1}^s \beta_{i-1}
(1-\beta_i \sigma^s )}}\eqno{(A)}$$
$$\
x_{i+1}^s = x_i^s -\beta_i (\zeta_{i+1}^s / \zeta_i^s ) p_i^s $$

$\}$

$$\eqalign{
r_{i+1}&=r_i +\beta_i w_i\cr
\rho_{i+1}&=(r_{i+1},r_{i+1})\cr
\alpha_{i+1} &= \rho_{i+1}/\rho_i}$$
$$
p_{i+1} = r_{i+1} +\alpha_{i+1} p_i
$$

{\bf \romannumeral2.} for $s=1,2,...n$:

$\{$
$$
p_{i+1}^s =\zeta_{i+1}^s r_{i+1} + 
\alpha_{i+1} (\zeta_{i+1}^s / \zeta_i^s)^2  p_i^s\eqno{(B)}
$$

$\}$

\noindent If $\sqrt{\rho_{i+1}} < {\rm tolerance}\cdot b_{\rm norm} $, exit.

\noindent $\}$

Let me make a few comments on the algorithm:
\noindent\item{$\bullet$} The denominator in line $(A)$ 
above will underflow for
larger $\sigma^s$ before convergence has been reached for the lower
$\sigma^s$. In practice the number of 
already converged $s$-values is stored 
and updated, so for any $s$ the CG algorithm is not  
executed past convergence.
\noindent\item{$\bullet$} Line $(B)$ differs slightly from the
explicit pseudo-code in [\multiplemass].
\noindent\item{$\bullet$} If one uses $H^2 +\sigma^1$ as driver in
$H^2$'s stead, $p$ can be replaced by $p^1$, resulting in some small
saving. 
\noindent\item{$\bullet$} The core conjugate gradient constitutes
of what is left of the above after the blocks
{\bf \romannumeral1} and {\bf \romannumeral2} are excluded.

We are only interested in obtaining a numerical approximation
to the vector $y=\epsilon(H)b$: The output we need is the 
linear combination
$$
y_{i+1}=\sum_{s=1}^n w^s x_{i+1}^s .\eqno{(3)}$$
In equation (3) I assumed that convergence of the 
CG procedure is achieved at step $i+1$ for all $s$. To avoid underflows,
one sometimes needs to replace $i$ by $i_s$ where $i_s$ is the $s$-dependent
point of convergence. I sketched above how this is done. 

It is obvious
that we are not interested in the individual vectors $x_{i+1}^s$, but
only in one particular linear combination. The idea is then to calculate
only the needed linear combination iteratively, without extra storage.
Just as in the Lanczos case, this requires an extra pass over the
core CG procedure.

Each $x^s_{i+1}$ in equation (3) is a linear combination
of the $s$-independent Krylov vectors $r_j$. Thus, $y_{i+1}$,
which is all we care to know, is also a linear combination of
$r_j$'s, and the single place $s$-dependence enters is in
their coefficients. 
One cannot compute only the linear combination 
of equation (3) in one pass: The contributions from Krylov vectors 
computed in the early stages of the iteration enter with coefficients
that are determined only in later iterations. 
Thus, a first pass is needed for the sole purpose
of computing the coefficients $\alpha_j$ and $\beta_j$ up to
the point where the driving CG process has converged. With this
information, one can calculate the coefficients $\zeta_j^s$ and
also the points of convergence for the different masses ($s$-values).
This information is used to compute all needed coefficients of 
the $s$-independent vectors $r_j$ in the decomposition of $y_{i+1}$.

The basic recursion determining the $x_j^s$ 
can be read off the algorithm listed above:
$$
\pmatrix {x_{j+1}^s \cr p_{j+1}^s } =
\pmatrix {1 & -\beta_j^s \cr 0& \alpha_{j+1}^s \cr}
\pmatrix {x_j^s \cr p_j^s } +\zeta_{j+1}^s 
\pmatrix {0\cr r_{j+1}}.\eqno{(4)}$$
In equation (4) $\beta_j^s=\beta_j \zeta_{j+1}^s/\zeta_j^s$ and
$\alpha_{j+1}^s = \alpha_{j+1} (\zeta_{j+1}^s /\zeta_j^s )^2$.
Starting from equation (4) I derived the following expression:
$$\eqalign{
y_{i+1} &=\sum_{k=0}^{i+1} R_k r_k\cr
R_k &= \cases{
         - \sum_{l=0}^{i-k} \left [ \beta_{k+l} \left (
           \prod_{j=1}^l \alpha_{k+j} \right )
          \sum_{s=1}^n w^s {{\zeta_{k+l+1}^s \zeta_{k+l}^s}
           \over {\zeta_k^s}} \right ],&for $i\ge k$;\cr
          \sum_{s=1}^n w^s \zeta_{i+1}^s,&for $k=i+1$.\cr}
\cr}\eqno{(5)}$$
In equation (5) I adopted the convention that $\left ( \prod_{j=1}^0 
\alpha_{k+j} \right )\equiv 1$.
Again, the upper bounds on the sums
over $s$ in equation (5) must be altered to prevent underflows, but the
information is available. So, in practice the sums over $s$ only
include the $s$-values that correspond to CG processes 
that have not yet converged at iteration $k$. 

Equation (5) clarifies why a single pass would not work without
the extra storage: $R_k$, for low $k$'s, depends on $\alpha_j$,
$\beta_j$, with $j$-values up to $i$. 
However, the coefficients $R_k, k=0,...,i+1$ can be computed after the
first pass. To compute $y_{i+1}$ we need the vectors $r_k$ and hence a
second pass. In the second pass no inner products are required,
since the coefficients $\alpha_j$ and $\beta_j$ are already known.

\vskip .5cm
{\bf Pseudocode for the two-pass version.}
\vskip .3cm
\noindent Input vector: $b$.

\noindent Vector variables: $w$, $r$, $p$, $x$.
 
\noindent Scalar variables and arrays: 
$\alpha_j$, $\beta_j$, $\zeta_j^s$, $\rho$, $R_k$, 
${\rm tolerance}$, $b_{\rm norm}$.
\vskip .1cm
{\bf 1. First pass.}
\vskip .2cm
\noindent{\it Initialize}: $r_0=b$, $p_0=b$, $\rho_0 =(b,b)$, $b_{\rm norm}=
\sqrt{\rho_0}$, $\alpha_0 =0$.
\vskip .2cm
\noindent{\it Iterate}: for $i=0,1,2,3,...$:
\vskip .1cm
\noindent $\{$
$$\eqalign{
w_i &= H^2 p_i \cr
\beta_i &=-\rho_i / (p_i, w_i )\cr}$$
$$\eqalign{
r_{i+1}&=r_i +\beta_i w_i\cr
\rho_{i+1}&=(r_{i+1},r_{i+1})\cr
\alpha_{i+1} &= \rho_{i+1}/\rho_i\cr
p_{i+1} &= r_{i+1} +\alpha_{i+1} p_i\cr}
$$
\noindent If $\sqrt{\rho_{i+1}} < {\rm tolerance}~ b_{\rm norm} $, exit.

\noindent $\}$
\vskip .2cm
{\bf 2. Compute} $\zeta_j^s$ from line (A) in the single pass
version, and $R_k$ from equation (5).
\vskip .2cm
{\bf 3. Second pass.}
\vskip .1cm
\noindent{\it Initialize}: $r_0=b$, $p_0=b$, $x_0=R_0 b$.
\vskip .1cm
\noindent{\it Iterate}: for $k=0,1,2,3,...i$:
\vskip .1cm
\noindent $\{$
$$\eqalign{
w_k &= H^2 p_k\cr
r_{k+1}&=r_k+\beta_k w_k\cr
p_{k+1} &= r_{k+1} +\alpha_{k+1} p_k\cr
x_{k+1} &=x_{k}+R_{k+1} r_{k+1}\cr}
$$
\noindent $\}$

For large enough lattices one might expect the second version to take
twice as long as the first. However, the nonuniform architecture of
the memory is crucial and the actual efficiency attainable in practice
can vary. I used a high level Fortran 90 code modeled on a
package described in [\rebbi] and ran
it at 64 bit precision 
on an SGI O2000 with four processors, each with 4MB cache memory.
All parallelizations were done using automatic options. I tested the methods
on a three dimensional system with gauge group $SU(2)$, two flavors [\prl], 
and lattice size $8^3$. I used gauge configurations generated from
the single plaquette gauge action with 
coupling $\beta=3.5$ and evaluated several
of the lowest eigenvalues of $DD^\dagger$. $n$ was set to 32. 
This computation is of the same order of magnitude as a few $D$-inversions.
I found that the second
method actually ran faster by $30\%$ than the first, at the same accuracy.
The observed speed-up instead of the expected slow-down most likely
reflects cache usage. Carefully optimized codes ought to show at least
some slow-down in the two-pass method. This slow-down will not
exceed a factor of 2.

The above test was done using the original method of [\prl], rather
than the refined version of [\scri]. In the original method,
the quantities $w_s$ and $\sigma_s$ are extracted from [\higham]:
$$
\epsilon(H)=\lim_{n\to\infty} {H\over n} \sum_{s=1}^n
{1\over{
H^2 \cos^2 {\pi\over{2n}} (s-{1\over 2}) +
\sin^2 {\pi\over{2n}} (s-{1\over 2})}}.\eqno{(6)}$$
In the refined version of [\scri] different weights $w^s$ and pole
locations $-\sigma^s$ are used.
The advantage of the refinement is claimed to be that similar
accuracies can be achieved with smaller values of $n$. This does
not come entirely for free though. In the original method the
spectrum of $\epsilon (H)$
is approached from within the interval $(-1,1)$ while in the
refined method of [\scri] the approach 
is oscillatory around $\pm 1$. Even tiny
excursions of the spectrum of $\gamma_5 \epsilon(H)$ 
below $-1$ can cause problems when one inverts
$D$. In applications one probably
would like to optimize for the accuracy of $1/D$ rather than
the accuracy of $\epsilon(H)$.
In the two-pass method the dependence of the overall computational
cost on $n$ has been sufficiently weakened to make the original
version of the rational approximation possibly more attractive.

In a recent paper [\borici] another method was presented, also 
requiring no extra storage. This new method employs a
Lanczos  iteration (similar to CG)
to tridiagonalize $H$. Next, the square root of the
relatively small triangular matrix is taken by
some spectral method (this is similar to old implementations
of the overlap [\twod]). To compute the action of the approximate
$\epsilon (H)$ on a vector $b$ a second pass over the
Lanczos procedure is needed. So, the ideas of using
two passes in order to avoid extra storage are similar.

It is premature at the moment to decide which method is best;
the answer may end up dependent on the particular 
architecture of the computer one uses. 
Therefore, it is useful to keep all possibilities in
mind, and avoid getting locked into one particular implementation.
New tricks producing large savings are still quite possible,
given the limited time people have been 
experimenting with implementations of $D$. 
In spite of the quite short time that has passed since
[\exactlymassless], looking back to 
the first implementations in [\twod], followed by [\twchiu]
where the Newton iteration was first applied, to more recent work
[\prl, \scri, \borici], I find the amount of progress
quite impressive.

{\bf Acknowledgments: } This work was supported in 
part by the DOE under 
grant \#DE-FG05-96ER40559.
I am grateful to Claudio Rebbi
for providing me with the 
qcdf90 package [\rebbi]
which I adapted to my needs in this work.
I am grateful to Artan Bori{\c c}i for communications
following the posting of [\borici] and for comments
following the posting of the first version of the present 
paper. I am indebted to
R. Edwards, Kei-Fei Liu and R. Narayanan for sharing
with me some of their experiences in implementing $D$. 

\vskip 2cm

\centerline{\bf References.}
\medskip 
\item{[\plbfirst]} R. Narayanan, H. Neuberger, 
Phys. Lett. B302 (1993) 62.
\item{[\npblong]} R. Narayanan, H. Neuberger, Nucl. Phys. B412 (1994) 574; 
Nucl. Phys. B443 (1995) 305.
\item{[\exactlymassless]} H. Neuberger, Phys. Lett. B417 (1998) 141;
Phys. Lett. B427 (1998) 353.
\item{[\trunca]} H. Neuberger, Phys. Rev. D57 (1998) 5417.
\item{[\prl]} H. Neuberger, Phys. Rev. Lett. 81 (1998) 4060.
\item{[\scri]} R. G. Edwards, U. M. Heller, R. Narayanan, hep-lat/9807017.
\item{[\multiplemass]} B. Jegerlehner, hep-lat/9612014.
\item{[\matrixbook]} G. H. Golub, C. F. Van Loan, 
Matrix Computations, John Hopkins University Press.
\item{[\lanczos]} J. Cullum, R. A. Willoughby, 
``Lanczos Algorithms for Large Symmetric Eigenvalue Computations'',
Vol. I, ``Theory'', Birkh{\" a}user, 1985.
\item{[\rebbi]} I. Dasgupta, A. R. Levi, V. Lubicz, 
C. Rebbi, Comp. Phys. Comm. 98 (1996) 365.
\item{[\higham]} Nicholas J. Higham, in 
Proceedings of ``Pure and Applied Linear Algebra:
The New Generation'', Pensacola, March 1993.
\item{[\borici]} A. Bori{\c c}i, hep-lat/9810064.
\item{[\twod]} R. Narayanan, H. Neuberger, P. Vranas, 
Phys. Lett. B353 (1995) 507.
\item{[\twchiu]} Ting-Wai Chiu, Phys. Rev. D58 (1998) 074511.

\vfill\eject 
\end